# Trade-Offs in Deploying Legal AI: Insights from a Public Opinion Study to Guide AI Risk Management

Kimon Kieslich[1,2] · Sophie Morosoli[1,3] · Nicholas Diakopoulos[4] · Natali Helberger[1,5]

**Abstract**

Generative AI tools are increasingly used for legal tasks, including legal research, drafting documents, and even for legal decision-making. As for other purposes, the use of GenAI in the legal domain comes with various risks and benefits that needs to be properly managed to ensure implementation in a way that serves public values and protect human rights. While the EU mandates risk assessment and audits before market introduction for some use cases (e.g., use by judges for administration of justice) other use cases do not fall under the AI Acts' high-risk classifications (e.g., use by citizens for legal consultation or drafting documents). Further, current risk management practices prioritize expert judgment on risk factor identification and prioritization without a corresponding legal requirement to consult with affected communities. Seeing the societal importance of the legal sector and the potentially transformative impact of GenAI in this sector, the acceptability and legitimacy of GenAI solutions also depends on public perceptions and a better understanding of the risks and benefits citizens associated with the use of AI in the legal sector. As a response, this papers presents data from a representative sample of German citizens (*n*=488) outlining citizens' perspectives on the use of GenAI for two legal tasks: legal consultation and legal mediation. Concretely, we i) systematically map risks and benefit factors for both legal tasks, ii) describe predictors that influence risk acceptance of the use of GenAI for those tasks, and iii) highlight emerging trade-off themes that citizens engage in when weighing up risk acceptability. Our results provides an empirical overview of citizens' concerns regarding risk management of GenAI for the legal domain, foregrounding critical themes that complement current risk assessment procedures.

## 1. Introduction

The widespread use of generative AI (GenAI) tools has left no societal domain untouched, including the legal sector. Reports and incidents show that GenAI is being widely adopted for various legal tasks, such as providing procedural assistance, offering legal consultations, and supporting judicial decision-making processes (Villasenor, 2024). Due to the widespread accessibility of GenAI tools, citizens can receive legal advice – or what resembles legal advice – at any time at minimal cost. Proponents of GenAI in the legal sector anticipate greater efficiency and increased access to justice, which could counteract social inequalities in accessing legal assistance (Chien & Kim, 2024). However, using GenAI for legal tasks is sensitive, as wrong or incomplete recommendations or decisions can have detrimental impacts on individuals and society, even contributing to a delegitimization of legal institutions as a whole (Helberger, 2025).

✉ Kimon Kieslich (corresponding author)
k.kieslich@uni-hohenheim.de

[1] AI, Media & Democracy Lab, University of Amsterdam, The Netherlands
[2] UKUDLA, University of Hohenheim, Germany
[3] Amsterdam School of Communication Research, University of Amsterdam, The Netherlands
[4] Communication Studies & Computer Science, Northwestern University, Evanston, United States
[5] Institute for Information Law, University of Amsterdam, The Netherlands

In light of the growing impact of AI on individuals and society, EU regulators have defined some uses of AI for legal purposes as high-risk applications in the EU AI Act. For instance, AI tools designed to be used by judges for the "administration of justice and democratic processes" (Annex 3, EU AI Act) must undergo proper testing and risk assessment before market introduction. This mandatory risk assessment is an important step in ensuring that AI aligns with human rights and societal values. However, questions remain about the effectiveness of these requirements and methods in fully capturing the potential risks (Kieslich et al., 2025; Orwat et al., 2024). Additionally, the administration of justice definition in the AI Act is a narrow scope that excludes use cases where lawyers or citizens use AI for legal purposes, such as informing themselves about the law or drafting legal documents. In these use cases risks might emerge that are currently under the radar of regulators (e.g., potential misinformation that is taken as a basis for a legal claim) but still might cause significant consequences for individuals or the functioning of the legal system (e.g. losing a legal case due to misinformation). Consequently, the current literature misses a systematic *public* evaluation of the use of AI in the legal sector, including an open mapping of risks and benefits factors regarding its use, as well as prevalent trade-offs that citizens engage in when evaluating the risk acceptance of the use of AI for legal tasks. Yet, these perspectives are of critical importance given that citizens are not only using, but also shaping the way GenAI is used for legal purposes, as well as the bias that expert assessors have in identifying and managing risks (Bonaccorsi et al., 2020; Reuel et al., 2025). Lived experiences of citizens are thus a crucial addition to fully uncover technology risks for the purpose of AI alignment to societal values (Kieslich, Diakopoulos, et al., 2024).

As a response, this paper presents survey results from a national representative sample of German citizens (N=488) to map and evaluate emerging risks and benefits of the application of AI in the legal sector, specifically for the tasks of legal consultation (e.g., using a Chatbot to answer legal questions), and legal mediations (e.g., using a GenAI tool to settle legal conflicts). We chose those tasks as these tasks refer to different stages of the legal process, involve different stakeholders, and come with different risk assessment obligations as legal consultation falls into the limited risk category whereas legal mediation falls into a high-risk category under the EU AI Act. Our results show that citizens are capable of uncovering a broad variety of risks and benefits and engage in complex trade-off deliberations regarding the use of AI in the legal domain. Consequently, this paper demonstrates the value of incorporating citizen input in risk assessment processes. It also implements a method to include citizens in trade-off decisions about the acceptability of particular risks from GenAI, and identifies prevalent trade-off themes that regulators and technology developers should pay attention to when introducing legal AI tools into society.

## 2. AI in the Legal Sector

The legal sector – like many other domains – faces challenges and changes due to the widespread use of AI. Surveys among legal professionals show that many practitioners already use, or plan to use AI systems for various tasks in the future (ELTA, 2023; Henry, 2024). These tasks can occur throughout the legal process, from legal research, administrative tasks, to assistance of drafting documents and legal decision-making (Laptev & Feyzrakhmanova, 2024). Empirical research shows that efficiency gains are seen as a major benefit from using AI for legal tasks for legal professionals (Chien & Kim, 2024; Schwarcz et al., 2025). It has also been argued that AI might increase access to justice for lay stakeholders, for instance due to the possibility to explain legal text, or to support filing notions (Chien et al., 2024). Judges also perceive benefits in terms of the automation of routine tasks (Dhungel & Beute, 2024).

Yet, scholars warn about the risks that AI systems pose for the legal sector. On the technical level, studies showed that AI tools produced unreliable – either wrong, or too vague – advice (Magesh et al., 2024). Importantly, the results differ considerably regarding the AI system used, with specialized legal systems performing better than open access systems (Ryan & Hardie, 2024). Findings like these question if potential accessibility gains are offset by the risk of inaccurate information provided to users. Further, and more fundamentally, scholars fear that the legal sector might face legitimacy problems due to the increased usage of AI (Helberger, 2025; Martinho, 2024; Socol De La Osa & Remolina, 2024) and point to concerns about the procedural rights of citizens (Dhungel, 2025; Metikos, 2024), i.e. concerns about a decrease of justice in terms of the quality of judgements in general. Helberger further points out several legitimacy threats for the legal sector: She argues that the institutions of courts face challenges due to a shift towards online conflict resolution rather than physical conflict resolution (e.g. in court), the mismatch between the development of uniform tools for a global market and the specific requirements of national or regional jurisdictions, and the erosion of legal expertise due to AI-powered tools that are optimised for goals like efficiency and speed instead of judicial values (Helberger, 2025). Another crucial challenge is due to the tendency of AI systems to generalize findings. Researchers fear that this could lead to an increase in societal bias as well as a lack of a human element, i.e. the potential of legal professionals to value and incorporate human emotions, conditions, and contexts in legal judgements (Martinho, 2024; Socol De La Osa & Remolina, 2024). Consequently, the use of AI in the legal sector entails risk-benefit trade-offs and warrants proper risk management approaches.

### 3. Risk Management of AI in the Legal Sector

In order to regulate the use of AI in the legal sector, many countries follow risk-based approaches (De Gregorio & Dunn, 2022; Ezeani et al., 2021). That means that (multiple) entities must be established to engage in risk management. According to Gellert, risk management include two related but distinct phases: risk assessment and risk management (Gellert, 2020). In the risk assessment phase, risk assessor needs to identify and map distinct risks, and assess their likelihood, severity, and magnitude, as well as estimate their consequences on individuals, organizations, and society. In other words: Risk assessment involves identifying potential risks, estimating the probability of occurrence, and also determining how severe those risks are for different entities. In a next step, the risk management phase, assessors need to decide on how much risk is acceptable and to what extent it can managed with adequate mitigation measures (Gellert, 2020). This step commonly involves risk-benefit trade-off decisions, where assessors are tasked to calibrate appropriate risk levels, as well as the definition of counter measures that aim to protect citizens. Practically, numerous academic, organizational, and governance frameworks have been established leading to a multitude of different risk classifications of AI (Slattery et al., 2024; Stahl et al., 2023). Yet, practical risk assessment can only be as good as the approaches that are applied in practice; and indeed researchers critique the conceptual vagueness of risk assessment (Moss et al., 2021). Therefore, it is crucial to interrogate the legal obligations for tech companies that aim to introduce AI for the legal sector, as well as the commonly used methods to assess and manage those risks. By examining the conceptual and empirical work in the literature on risk and impact assessments, we identified several shortcomings in current regulatory approaches and risk assessment practices.

Structurally, risk assessment and management processes often lack democratic accountability in prioritizing the vested interests of technology companies (Moss et al., 2021). The EU places the responsibility of conducting risk assessments primarily on companies themselves (Griffin, 2024; Orwat et al., 2024), which introduces biases by assigning the responsibility to interested parties. This power asymmetry is also mirrored in more normative questions like the decisions on value conflicts, which level of various risks might be acceptable for society but also, and importantly, who is entitled to take that decision (Gillespie, 2024). In other words, interested companies have typically considerable discretion to organise the internal process of risk management, making this process vulnerable to capture, a conscious or unconscious avoidance of fundamental value questions and a lack of external legitimacy (Orwat et al., 2024).

Further, risk management is heavily expert focused, while largely neglecting laypeople's input (Ebers, 2024). This is concerning because it has been shown that experts' opinions are limited and often ignore the lived experiences of affected individuals (van der Heijden, 2021). Similarly, risk assessors often need to decide on risk-benefit trade-offs, i.e., how much risk is acceptable in relation to the potential benefit – a decision that is largely dependent on the assessors themselves (Schmitz et al., 2024). This decision-making process is far from neutral because the judgment is dependent on societal, cultural, commercial and individual factors (Gillespie, 2024; Orwat et al., 2024). It also challenges the democratic legitimacy of AI-assisted legal and judicial proceedings if citizens and society have no role whatsoever in deciding which uses of AI in the legal domain are acceptable or not. This is further corroborated by the critique that many established risk metrics reduce the complexity of data or information, resulting in the loss of important contextual information (Gellert, 2020). Therefore, scholars argue for supplementing risk assessment toolboxes with more qualitative, and context-aware approaches (Walker et al., 2024), including approaches that capture citizen's concerns and expectations in trade-off decisions.

Further, scholars have argued that risk management is heavily focused on technical and quantifiable risks rather than societal, or individual risks. While it is comparatively easy to calculate AI related errors, it is far more complicated to identify and assess risks for fundamental rights (Kieslich et al., 2025; Orwat et al., 2024). But especially for high-impact domains like the legal sector, these issues are of high democratic importance. For instance, faulty AI systems in the legal domain might lead to wrongful convictions, which leads not only to material consequences and emotional damage for affected parties but also harm public perceptions of the legitimacy of justice as an institution. Yet, these harms are highly subjective and are routed in contextual and personal experiences, which are outside the scope of most technology developers or regulators. In response to the critique, we argue, as do other scholars, that the voices of affected citizens and communities need to be more strongly included in risk assessment and management approaches (Matias & Price, 2025). Public perceptions are especially important given the use case of AI in the legal sector. Ultimately, citizens are on the receiving end of changes in the legal system, and they might not have a choice to engage with an AI system or not. For instance, in a legal conflict one party could decide to use AI for legal consultation, and the other party needs to react to its use. Further down the line, one can also imagine the use of AI to settle legal conflicts. Yet public input is currently not necessitated by European (and German) law (Ebers, 2024). In the following, we give an overview on the scholarly literature on public opinion of AI (risks) and position our approach in the existing literature.

## 4. Public Opinion for Risk Management of AI

In light of the rising importance of AI for individual and societal life, scholars have invested in measuring public opinion on AI; including literature review approaches (Dreksler et al., 2025; Eom et al., 2024) and projects that follow longitudinal approaches to track public opinion over time (AlgoSoc, 2025; Kieslich et al., 2023; Zhang & Dafoe, 2020). Public opinion on AI is deemed relevant especially for those cases where it impacts the public interest, i.e. when the use of AI affects the individual life of all citizens (Züger & Asghari, 2023). Rahwan proposed a society-in-the-loop (SITL) approach that argues for the measurement and acknowledgement of societal values when developing AI systems (Rahwan, 2018).

Especially important for the purpose of this study is the line of work that is done on risk perception of AI, and the trade-offs between perceived or anticipated benefits and risks (Araujo et al., 2020; Brauner et al., 2025; Kieslich et al., 2022; Klein et al., 2024; Said et al., 2023; Schwesig et al., 2023; Wei et al., 2025; Wilczek et al., 2024). Studies have shown that heightened risk perceptions are correlated with hesitancy to adopt AI technologies as well as the preference for stronger regulations (Wei et al., 2025; Wilczek et al., 2024). In addition, researchers also found that those with higher knowledge about AI tend to underestimate AI-related risks (Said et al., 2023). A national representative study of the German population measured risk and benefit perceptions of 71 different AI tasks (Brauner et al., 2025). The results show that risk-benefit trade-offs are context dependent, and that legal applications like *deciding on legal cases* are perceived as high-risk applications that yield only limited societal benefits (Brauner et al., 2025). These findings are in line with nationally representative studies from Germany and the Netherlands that report that especially the uses of AI in the legal system are deemed as particularly risky and thus have low approval among the respective populations (AlgoSoc, 2025; Araujo et al., 2020; Kieslich et al., 2021).

Yet, most studies measuring risk perceptions focus largely on a generalized perception of risks (e.g., agreement scales from not risky to risky or similar) but fail to measure what exactly it is that constitutes those risk perceptions. In other words, we do know i) how much risk is seen for *specific* application scenarios within domains, ii) what socio-demographic, individual or societal factors influence those risks perceptions, iii) what effects risk perceptions of AI have in regard to technology adoption and regulatory preferences and iv) what considerations matter to citizens when deciding whether a particular risk is societally acceptable or not. For instance, Brauner et al. (2025) measure risk perceptions of the AI task *deciding on legal cases*. While these generalized risk perception ratings are an important information, they merely scratch the surface as they don't provide further information what these risk perceptions are based on.[1] One can imagine a wide range of potential risk factors, spanning from more technical concerns like AI-related errors that lead to wrongful convictions, to bias in AI systems that lead to discrimination in legal judgments, to more societal concerns like a lack of empathy in legal decision-making. We argue that mapping and measuring these detailed concerns but also the associated trade-off decisions add an important layer to the studies of public risk perceptions of AI. That is, knowing *specific* risk factors can inform potential risk management steps in showing what risk factors need to be especially acknowledged when proposing and implementing regulatory mechanisms.

---
[1] This is not a critique at the study of Brauner et al. (2025) since their research interest was to compare various AI tasks with each other.

In our study, we measure perceptions of the use of generative AI in the legal sector using a narrative scenario. Scenarios are often-used to elicit reactions in public opinion studies (Said et al., 2023), and are especially useful in showcasing lived experiences (Matias & Price, 2025). We focus on two tasks where AI is expected to alter legal processes considerably: legal consultation and legal mediation. Importantly, both tasks can potentially be accessed by citizens that face legal conflicts. Legal consultation means that users utilize GenAI tools to inform themselves on the legal case. Here, users can describe their legal conflict, ask the AI for a legal opinion and legal options, and also support or automatically draft legal documents (e.g., complaints). Legal mediation describes the use of GenAI in legal decision-making, i.e. settling conflicts. Here, a GenAI tool instead of a human judge, processes the legal documents from conflict parties and proposes a legal solution (e.g., settlement). The validity of this mediation use might differ from binding decisions to non-binding recommendations for conflict parties. While the legal consultation task is classified as non-high risk under the AI Act, the legal mediation tasks falls under the high risk category of the EU AI Act. Yet, both legal tasks entail risks and benefits. However, we know little about public perceptions of associated risks and benefits (what risk and benefit factor are seen) as well as the overall evaluation of these trade-offs.

Thus, we propose the following research questions:

> *RQ1: What specific risks and benefits do citizens anticipate from the use of GenAI for legal consultations and legal mediation?*

Further, to embed our study in the broad literature on public opinion on AI, we follow a mixed methods approach that provides quantitative findings elaborating on predictors for risk acceptance of AI for legal tasks. Here, we extend the existing literature on public opinion in including the factors subjective knowledge of AI (Cave et al., 2019), personal use of AI technology (Dreksler et al., 2025), experienced discrimination (Kieslich & Lünich, 2024; Mun, Au Yeong, et al., 2025), political position (Bao et al., 2022; Mun, Au Yeong, et al., 2025), socio-demographic factors (Bao et al., 2022; Kieslich et al., 2022; Mun, Au Yeong, et al., 2025), as well as fairness perceptions of AI and the legal system (Starke et al., 2022). While we know a lot about general predictors of risk acceptance of AI (AlgoSoc, 2025; Brauner et al., 2025), specific research on predictors of risk acceptance in the legal domain remains scarce. Thus, we pose the following research questions:

> *RQ2: What socio-demographic, political, and AI-related factors influence acceptance of risk for GenAI in legal consultations and legal mediation?*

Lastly, we are not only interested in the prevalence of risk and benefit factors (RQ1) and the quantified risk-benefit trade-off perception (RQ2), but also in individual trade-off considerations that citizens engage in when deciding on risk acceptance. To be precise, we are not only interested in which risks and benefits are anticipated, but also how they are weighted against each other when making a judgment about whether or not to accept a risk. For instance, respondents can raise concerns on an increase in biases through the use of GenAI in legal decision making, but can simultaneously see benefits due to an increase in access to justice as legal consultation might get cheaper. Asking for the trade-off process can then inform research on which arguments are weighed more heavily, and why. As there is limited guidance for trade-off decisions for AI practitioners and regulators in practice due to the lack of public input and knowledge about public perceptions regarding those trade-offs, highlighting prevalent risk-benefit trade-off considerations are important data points for risk management researchers, regulators and practitioners alike. Consequently, we ask the following:

> *RQ3: What risk-benefit trade-offs emerge when considering the acceptance of risk for GenAI in legal consultations and legal mediation?*

## 5. Method

### 5.1. Procedure

To answer these research questions, we conducted a survey of German respondents from June 10 to June 24, 2025. We chose to focus on Germany because of its central role in European markets and its influence on EU politics. Respondents were recruited through an open access panel of the ISO-certified (ISO 20252:2019) market research institute Bilendi. We applied quotas for age groups, gender, and educational level to ensure the sample was representative of the German public with respect to these characteristics. We received ethics approval from the institutional ethics board of the first authors' affiliation.

After giving their consent, the respondents answered questions about their sociodemographic characteristics and political views, followed by questions about their perceptions of AI. Then, they read a fictional scenario about how GenAI could be used in the near future to resolve legal disputes. Writing and evaluating scenarios has proven to be a viable method for invoking brainstorming activities around technological developments, especially among non-expert respondents (Burnam-Fink, 2015). The selected scenario describes a conflict between a tenant and a landlord, in which both parties use GenAI for legal consultation purposes, i.e., to inform themselves about their rights. However, the parties fail to reach an agreement, so they go to court. There, another GenAI tool is used to mediate the conflict without the involvement of a human judge. The story is intentionally open-ended to elicit positive and negative reactions to the use of GenAI for these purposes (see Appendix 1 for the full text). The landlord vs tenant scenario, thus, describes a tangible use case that might have material consequences (e.g., loss of money) and is relatable for many German citizens, albeit not as high-stakes as some other scenarios might be.

After reading the stimulus, respondents provided open-ended answers for both use cases, detailing their perceptions of the risks and benefits of using GenAI for (i) legal consultation and (ii) legal mediation. They were then asked to rate their risk acceptance for both tasks on 5-point Liker scales ranging from negative ratings (risks outweigh the benefits) to positive ratings (benefits outweigh the risks) and elaborate in an open-ended text field on their reasoning for the rating, detailing their thoughts on the trade-off. Lastly, respondents evaluated the scenario in terms of plausibility, severity and magnitude on 5-point Likert scales before being debriefed. The survey took participants a median of 15 minutes to complete.

A total of 858 respondents began the survey after passing the quota check. We applied strict quality checks to ensure a high-quality sample. We filtered out respondents who 1) did not finish the questionnaire (*n*=259), 2) failed the attention check[2] (*n*=93), 3) completed the questionnaire in under five minutes (*n*=7), and 4) provided nonsensical answers to the open-ended questions

---

[2] For the attention check, we asked which legal conflict the scenario described (answer options: assault, traffic offence, rental dispute, divorce). Participants who answered *rental dispute* passed the attention check.

(*n*=9). Ultimately, our sample consisted of 488 respondents. The final sample consists of 235 men (48.1%), 244 (50.0%) women, and 9 non-binary persons or persons who don't want to assign themselves to one category (1.9%)[3]. The average age was 49.58 years (*SD*=15.64), and 132 (27.1%) respondents with low, 169 (34.6%) respondents with medium, and 187 (38.3%) respondents with high educational degree[4].

## 5.2. Measurement
### 5.2.1. Statements Ratings

*Risk Acceptance*. Risk acceptance was measured with one self-developed item for each of the legal tasks on a 5 point Likert scale (1=do not agree; 5=fully agree). The items read as follows: "Do you agree with the following statement? The benefits of using AI for [1) legal consultation; 2) legal mediation] outweigh the risks?" ($M_C$=3.29, $SD_C$=1.13; $M_M$=3.28 $SD_M$=1.15)[5]. Thus, high values indicate a greater benefit perception.

*Perceived Fairness of AI*. Perceived fairness of AI was measured via three items on a 5 point Likert scale (1 = do not agree; 5 = fully agree) that reads as follows: "Recommendations made by AI systems are trustworthy", "Recommendations made by AI systems are generally fair", and "Recommendations made by an AI are generally unbiased". We calculated a mean index for perceived fairness of AI, *M*=2.92, *SD*=0.97, Cronbachs *α*=.87. The item wording was self-developed but the categories adopted from theories of organizational justice (Colquitt, 2001).

*Perceived Fairness of the Legal System*. The measurement of perceived fairness of the legal systems was adapted to the one of perceived fairness of AI. Respondents answered the question "To what extent do you agree with the following statements about the state of the legal system in Germany?" on three items: "In Germany, legal disputes are generally resolved fairly", "In general, legal proceedings in Germany are unbiased", and "The German legal system is trustworthy" on a 5 point Likert scale (1 = do not agree; 5 = fully agree). We calculated a mean index from the three variables, *M*=3.20, *SD*=1.10, Cronbachs *α*=.91.

*Experienced Discrimination*. Experienced discrimination was measured with five items on a five point Likert scale (1 = never, 2= seldom, 3=sometimes, 4=often, 5=very often). Respondents were asked how often they experienced the following in daily life: Someone acts as if they are afraid of you; You receive worse service than other people in restaurants or stores.; You are threatened or harassed.; Someone acts as if you are not taken seriously.; You are treated with less respect than other people. We calculated a mean index over the five items, *M*=1.93, *SD*=0.73, Cronbachs *α*=.84. The measurement was adopted from Williams et al. (1997).

*Subjective Knowledge on AI*. Subjective Knowledge of AI was measured as a single item to the question "How would you rate your own knowledge of artificial intelligence in general?" on a five point Likert scale (1 = I know nothing about AI; 2 = I know a little bit about AI; 3 = I know a bit about AI; 4 = I know a fair bit about AI, 5 = I know a lot about AI), *M*=2.93, *SD*=0.91. The item wording was adopted from Došenović et al. (2022).

---

[3] For statistical reasons, we merged the data from non-binary and people who preferred no self-describe with the women category. We do so only for the analysis of RQ2. While we are aware of the ethical discussion about mering gender categories, we decided to do so as 1) we did not want to drop the answers from non-binary or people with other gender expressions from our analysis, and 2) we merge them with the woman category as it has been shown that AI can exert gender-biases that targets predominantly women, or people with other sexualities.

[4] Low level of education refers to lower secondary degree or no degree, medium level of education refers to upper secondary degree plus post-secondary non-tertiary degree, and high level of education refers to short-cycle tertiary, Bachelor, Master, and Doctoral degree.

[5] Subscripts C refers to legal consultation items and subscripts M to legal mediation items.

*GenAI use*. GenAI use was measured with a single item based on Kieslich et al. (2024). The item wording was "How often do you use generative AI tools? With generative AI tools we mean technology that can create new content (e.g. text, images, audio, video) based on input instructions, indexed material, and training data. Examples of generative AI tools are ChatGPT, Midjourney, Dall-E, Microsoft Copilot, and Claude". The item points were 1=not at all, 2=once a month, 3=two or three times a month, 4=once a week, 5=several times a week, and 6=daily, *M*=2.93, *SD*=0.91.

*Scenario Evaluation*. To check the adequacy of the stimulus, we measured *plausibility* of the scenario, as well as *severity* and *magnitude* of the described risks. Plausibility is a common quality criterion in the scenario-writing literature (Ramírez & Selin, 2014; Selin, 2006), whereas severity and magnitude of risks are crucial factors in risks management (European Data Protection Supervisor, 2025; Meßmer & Degeling, 2023). All items were measured on five point Likert scales. *Severity* was defined as a very serious or unpleasant condition. *Plausibility* describes how likely we think a scenario is to become reality. For magnitude, we posed the following question: "This scenario is an example that could happen to any number of people worldwide. How many people do you think could be affected by the damage described in this scenario?". All items were measured on five-point Liker scales (1=not severe/not plausible/a small number of people, 5=extremely severe / extremely plausible / a majority of people in society).

*Political Position*. Political position was measured on a 9-point Liker scale with lower values indicating a left political leaning and higher values indicating a right political leaning (*M*=4.68, *SD*=1.64).

### 5.2.2. Open Answers

*Risk and Benefits Mapping*. To assess the breadth of potential risk and benefits from the use of GenAI for both legal tasks, we asked respondents to elaborate in an open answer format on their a) risk and b) benefits perceptions for the use of AI for a) legal consultations, and b) legal mediation. Specifically, we asked four questions: "When you think about the scenario you have just read, where do you see [1) risks; 2) benefits] for individuals, organisations, specific social groups and/or society as a whole in the described use of AI for [1) legal consultation; 2) legal mediation]?"

We conducted a quantitative content analysis of the risks and benefits identified in the open-ended responses for both use cases. In a first step, two authors (native German speakers) each created an individual codebook using an exploratory approach by reading a random sample of 100 responses to each open question. In a second step, the two authors created a consensus codebook based on the individual codebooks (Helberger et al., 2020). The developed codebook including a description of the categories are supplemented in Appendix 2. Note that open answers could contain multiple risks/benefit mentions. For every open answer, we coded up to three risks and benefits. The authors conducted two pre-tests until the reliability test showed satisfactorily results (Cohens *κ*=.63 for Consultation Risks; Cohens κ=.72 for Consultation

Benefits; Cohens κ=.73 for Mediation Risks; Cohens κ=.82 for Mediation Benefits)[6]. Statements which were difficult to code were discussed in the team and coded together.

*Explanations of Trade-Off Decisions*. For getting deeper insights on the trade-off decision process of the articulated risks and benefits, we asked respondents to elaborate on their reasoning. Specifically, we asked: "Please explain why you believe that the risks described outweigh or do not outweigh the advantages of using AI for [1) legal consultation; 2) mediation of legal disputes]." After reading 100 responses for each question, the authors concluded that the breadth of the answers was too wide to develop a feasible codebook for a quantitative content analysis. Thus, the authors decided to perform qualitative thematic analysis of the answers and supplement the quantitative and qualitative findings with quotes from the answers (Glaser & Strauss, 2017; Lofland et al., 2022). For that, we highlighted and excerpted quotes from the open answers. Based on this selection, we identified emergent themes and re-structured quotes along these axes. This process was iteratively structured, so the material was scanned multiple times and constant comparisons were applied.

## 6. Findings

Respondents rated the depicted risks in the scenario as medium in severity ($M$=3.02, $SD$=1.11) and relatively high in magnitude ($M$=3.62, $SD$=1.02). This indicates that the risks are likely to affect many people in society and pose a tangible yet non-devastating risk. The scenario was also rated as highly plausible ($M$=3.68, $SD$=1.09), with only 51 respondents ranking it as not plausible, i.e. rating 1 or 2 on the 5-point Likert scale. Consequently, the scenario is a good stimulus because it describes a plausible future use of GenAI for legal conflict resolution that depicts a tangible risk that is tangible and leaves ample room for discussion about the acceptance of using GenAI for the depicted tasks.

In the following we describe the findings of our study. First, we map the risk and benefit factors to identify potential issues that may arise when using AI for legal tasks. Secondly, we present the results of an OLS regression analysis on risk acceptance. This provides more detailed information about risk acceptance for both types of legal task, showing which predictors lead to a more positive or negative attitude. Finally, we present a qualitative analysis of the trade-off explanations that respondents provided to justify their risk acceptance ratings.

### 6.1. Mapping of Risk and Benefit Factors

We identified a multitude of different risks and benefits for the use of GenAI for 1) legal consultation, and 2) mediation of legal conflicts. Figure 1 and 2 show the anticipated risks and benefits when using GenAI for legal conflict resolution. Figure 3 and 4 show the anticipated risk and benefits for the use of GenAI for the mediation of legal conflicts. Appendix 2a-d provides detailed descriptions of all categories.

#### 6.1.1. Risk and Benefit Factors for Legal Consultation

The results show that there is a wider breadth of anticipated risks than there is for benefits. That is, we identified 17 different risks, but only 8 different benefits. This indicates that benefit anticipation of GenAI for legal consultation clustered around a smaller number of issues, while risk perceptions were more diffuse.

---

[6] Note that the consultation risk category contained 20 coding options. Thus, achieving high reliability results is more difficult than for variables with fewer categories.

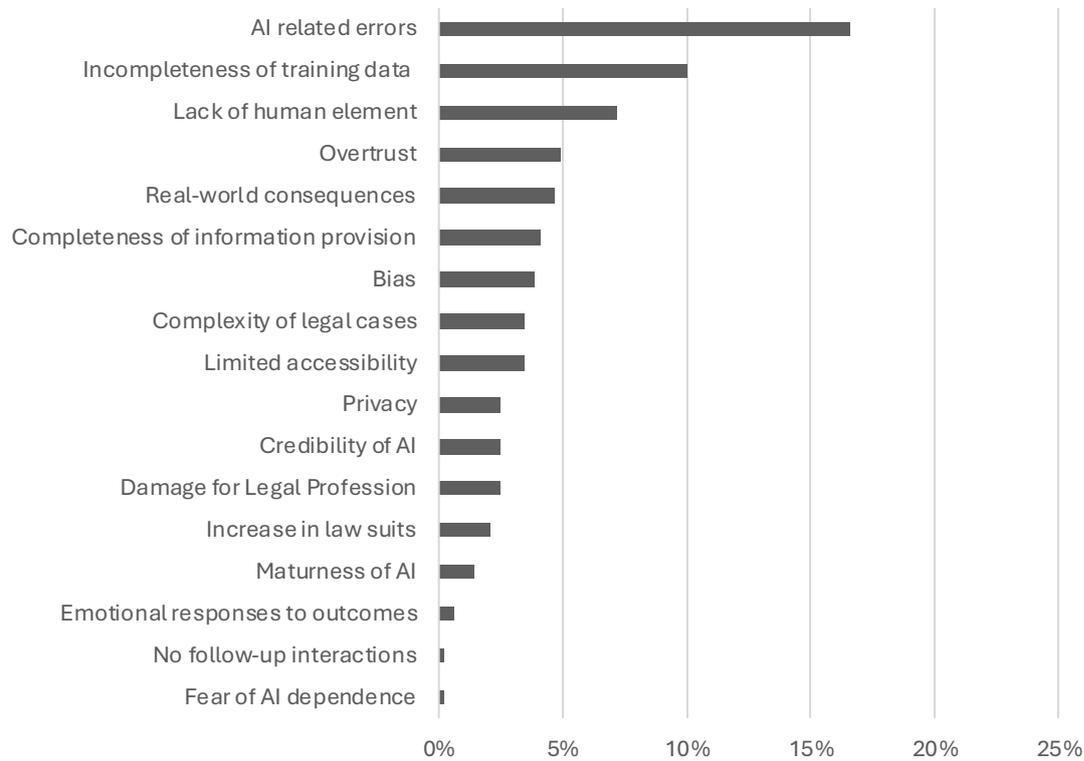

*Figure 1: Occurrence of Risk Factors (Legal Consultation)*

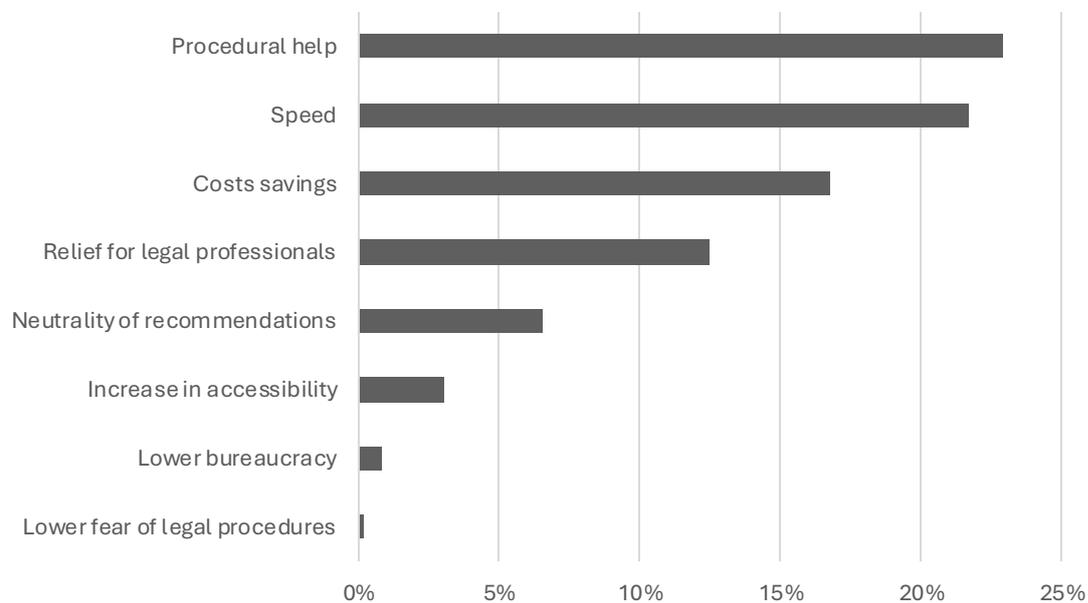

*Figure 2: Occurrence of Benefit Factors (Legal Consultation)*

The most common risks are connected to technical limitations, that is AI errors (e.g., AI makes mistakes or its output is too vague), followed by concerns about the data / knowledge base of the systems (e.g., AI does not know applicable laws). Yet, the third most common answer concerned the lack of a human element, i.e. the notion that humans are important in the evaluation of legal conflicts. This encompasses mentions like a lack of empathy, or missing humanness in recommendations. Further, there are several other risks that were only mentioned in under 5 percent of the answers. Yet, these categories describe important and original concerns like the question of validity of AI, i.e. if legal professionals / courts would take AI advice seriously, the completeness of information provision, i.e. concerns that users might manipulate the legal advice due to one-sided input of information / description of the situation, or societal concerns about an increase in law suits due to the ease of drafting documents.

On the benefit side, respondents anticipated the most advantages for procedural help, which encompasses support in drafting documents, documenting cases, or writing letters, but also includes the expectation that AI can give a good and comprehensive overview about legal options. Related to this, respondents anticipated time saving as second most prevalent benefit, followed by cost savings. Furthermore, 10 percent of respondents anticipated a relief of legal professionals that would help make the legal system more efficient. Quality improvements due to the help of AI was merely articulated in the hope that AI systems would come to more just and neutral decisions. Yet, in comparison to efficiency gains, the quality improvement played only a secondary role.

### 6.1.2. Risk and Benefit Factors for Legal Mediation

Again, the breadth of anticipated risks outweigh the number of anticipated benefits. We also identified a vast overlap in coding categories, yet, some differences emerge, which lead to the addition of new codes and the exclusion of others. For instance, we added emotional relief as a new benefit category, which refers to perceived lower emotional pressure due to the avoidance of a court case with a human judge.

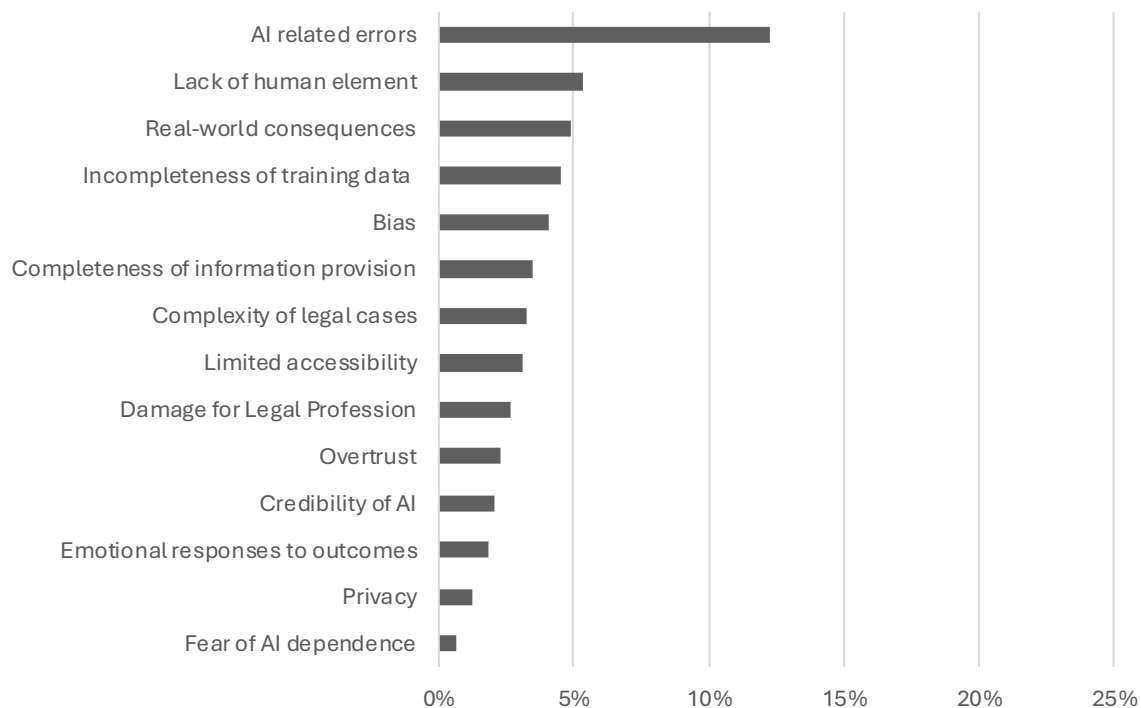

*Figure 3: Occurrence of Risk Factors (Legal Mediation)*

Similar to the anticipated risks for legal consultation, AI related errors is by far the most prevalent risk category. Interestingly, this is followed by the lack of a human element as second most prevalent risk indicating stronger individual/societal concerns than technical concerns for the legal mediation application.

On the benefit sides, we also observe differences in comparison to the legal consultation task. Here, time savings were seen as the primary benefit, followed by cost benefits. Procedural assistance, i.e. helping to get a comprehensive overview on options, was a less prevalent benefit than for legal consultation. Again, quality improvements, i.e. an increase in neutrality of decisions, was only a secondary benefit.

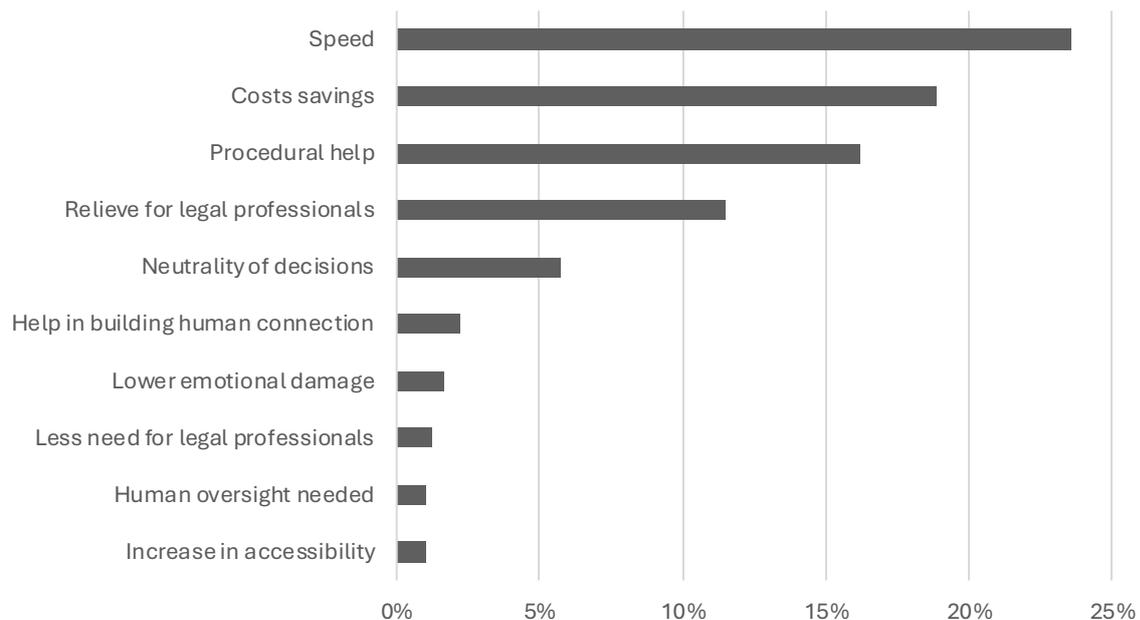

*Figure 4: Occurrence of Benefit Factors (Legal Mediation)*

### 6.2. Predictors for Risk Acceptance

In a next step, we investigated which socio-demographic, political, and AI-related factors influence acceptance of risk for GenAI in legal consultations and legal mediation (RQ2). For that, we calculated two separate OLS-regressions on the dependent variable risk acceptance for the legal consultation (model 1) and legal mediation (model 2) use case. The results are summarised in Table 1.

The results show that perception of AI fairness positively impacts risk acceptance for both use cases of GenAI in the legal field. That is, if people believe that AI systems can make fair recommendations, they tend to perceive that the benefits of AI outweigh the risks. On the contrary, fairness perceptions of the legal system in Germany did not impact the risk-benefit trade off perception for the use of GenAI for both tasks. Thus, risk acceptance of GenAI in the legal domain is independent from perceptions of how well the legal system works, i.e. for risk acceptance of GenAI in the legal domain it does not matter if people perceive the (current) legal system as fair or not. Moreover, we did not find significant effects for other AI-related variables like subjective knowledge about AI and usage behaviour of GenAI. Further, we found some differences in significant predictors for both models. For the legal consultation task, we found associations of risk acceptance with age, political position and education. Older, more right leaning, and people who hold a medium educational degree tend to perceive more that benefits outweigh the risks. For the legal mediation case, we found high education as a positive predictor, i.e. people who hold a high education degree are more convinced that the benefits of AI for legal mediation outweigh the risks.

|  | Model 1: Legal Consultation | | | Model 2: Legal Mediation | | |
|---|---|---|---|---|---|---|
|  | b | SE | β | b | SE | β |
| Intercept | 0.867 | 0.402 |  | 1.47 | 0.410 |  |
| Perception of AI Fairness | 0.272 | 0.060 | **.232\*\*** | 0.314 | 0.061 | **.263\*\*** |
| Perception of Legal System Fairness | 0.057 | 0.050 | .056 | 0.047 | 0.051 | .045 |
| AI Knowledge | 0.119 | 0.070 | .096 | 0.009 | 0.071 | .007 |
| Generative AI Usage | 0.038 | 0.041 | .055 | 0.070 | 0.042 | .101 |
| Pol. Position | 0.064 | 0.031 | **.093\*** | 0.038 | 0.031 | .054 |
| Age | 0.008 | 0.004 | **.107\*** | 0.005 | 0.004 | .075 |
| Gender (1 = male) | -0.022 | 0.101 | -.010 | 0.018 | 0.103 | .008 |
| Education (Medium) | 0.338 | 0.128 | **.143\*\*** | 0.180 | 0.130 | .075 |
| Education (High) | 0.204 | 0.133 | .088 | 0.270 | 0.135 | **.114\*** |
| Experienced Discrimination | 0.055 | 0.074 | .036 | -0.050 | 0.075 | -.031 |

Note:
adj. $R^2_{Consult}$ = .111, , *p<.05, **p<.01
adj. $R^2_{Mediation}$ = .110, *p<.05, **p<.01

*Table 1: Regression on Risk Acceptance*

### 6.3. Trade-Off Explanations

The open-ended answers[7] from the trade-off explanation question[8] showcase citizens' trade-off reasoning process of risks and benefits of the use of GenAI for legal consultation and mediation (RQ3). We used thematic analysis to identify trade-off themes from the answers. As these are similar for both use cases, we report on them jointly.

*The degree of human involvement*

A central theme in the explanation was the role of the human, especially regarding human capabilities, limitations of machines, and the structural role that participants thought humans should play in the future. Though, the role of the human is not agreed upon, as the open answers indicate a range from strong preference of human, over shared roles, towards favouritism of AI. Thus, some respondents expressed their opposition in firm words: "People are much better at resolving conflicts personally, and just as quickly. Human communication and activity will always beat AI." [R343]. Respondents miss the human judgment of cases: "Human judgment in each specific case will not always be guaranteed by AI." [R472], and the lack of emotions in particular: "The risks outweigh the benefits because AI cannot fully grasp the individual, emotional, and social aspects of a conflict." [R372]. Moreover, respondents stress the importance of the experience of judges: "I think it could be possible that the AI does not have all the legal basis. And that a judge has seen other cases and might act differently." [R19]. Some respondents also go a step further and deliberate about the societal impacts for an ever increasing impact of AI: "AI does not have ethical or moral thinking. I am against us giving up more and more and allowing our abilities to atrophy. Furthermore, it is always possible that AI could be working with misinformation." [R104]

---

[7] To identify the quotes within the data set, we added the respondent number in the format RNNN to the quotes. Quotes were translated with the automated translation model of DeepL (https://www.deepl.com/en). All of the translated quotes referenced in this section were double-checked by two native German-speaking authors.
[8] Question: Please explain why you believe that the risks described outweigh or do not outweigh the advantages of using AI for [1) legal consultation; 2) mediation of legal disputes].

Some respondents acknowledge human competencies, but weigh them off against potential benefits of AI. For instance, the accessibility of information against the value of human thinking: "The AI compiles the information. This means that previous court decisions and experiences are also incorporated into this advice. [...] AI cannot replace human thinking, but it can support us in our thinking." [R20].

For some respondents the risk acceptance depends on the level of human oversight. Respondents argue for the inclusion of lawyers, ("You would have to train artificial intelligence to do this, and lawyers would also have to be sitting in the background." [R288]), the clear definition of human domains vs. AI assisted domains ("AI can assess the facts comprehensively and expertly and help to reach an out-of-court settlement. But in court proceedings, it is ultimately the judge who must decide." [R341]), or the importance of double checking AI generated advice ("AI can be a valuable assistant, providing information and writing complicated texts. However, it should be checked by a human being beforehand." [R409]). Some respondents also outline more broadly the need for regulation to mitigate risks ("AI helps resolve conflicts faster and more cost-effectively. With clear rules, risks can be controlled effectively." [R305]).

*Neutrality vs Manipulation*

A contested notion that appeared throughout the answers was the degree of neutrality AI could exert. Some respondents indicated a great belief in this as the following answers indicate: "Among other things, I also see advantages in the incorruptibility and safe use of our legal system without having to tolerate influence or even misinterpretations by individuals." [R279]. Respondents point out the available knowledge base of AI systems, i.e. the information the AI was trained on, as a factor that fuel their belief in the quality of the output: "AI takes much more into account than a judge could. It could therefore be fair, save costs, and allow judges to devote themselves to more important cases." [R289]

Other respondents generally believe in the potential neutrality of AI, but also raise concerns regarding the misuse potential: "It is advantageous that AI takes a neutral position. However, AI is only able to access information that is provided to it. This can quickly lead to abuse if, for example, fake information is spread." [R257]. Moreover, respondents anchored their reasoning in their personal experiences with AI systems to challenge the capacity of GenAI to generate impartial output. This is particularly pronounced given GenAI's documented tendency to (overly) align its output with the user's expectations or beliefs, a phenomenon referred to as AI sycophancy: "AI is not yet at a level where it can really help in such matters, because it is easily manipulated. If you write to it that something is wrong, it often adopts this opinion. But in court, both sides count—the judge and the parties involved decide together based on their personal assessments." [R389]. That also leads to concerns about the general applicability for AI for legal conflict resolution as explained by the following quotes: "It is impossible to generalize, as the AI is confronted with two different opinions. The AI receives no information about emotions and does not know whether someone is lying." [R476] or "There are individuals who rely too heavily on the statements made by AI, even if these may not be legally binding. These individuals will then insist on false statements and information." [R41].

*Access vs Accessibility*

Another frequently mentioned trade-off theme evolved around the potential for wider access to justice versus differing levels of proficiency using AI tools which lead to a different utility of GenAI for population groups. Along the line of AI proponents, respondents see potential for greater access to justice due to interpretative help, time and costs aspects. Respondents put that as follows: "Information about legal situations is provided online in a way that is understandable for every user, which can only be a good thing." [R300] for the procedural help aspect, "Dealing with law and legal advice is very time-consuming due to the large number of laws and different source texts. In addition, legal texts are usually difficult to understand. AI can bridge this gap." [R384] for the time aspect, or "The advantages outweigh the disadvantages because AI facilitates legal access for people who cannot afford a lawyer" [R132] for the cost aspect.

Some respondents were more nuanced in their judgment of the risk acceptance, highlighting conditional risk acceptance. For instance, a respondent outlined the benefit for early legal consultation as a benefit "Proper legal advice is probably more accurate and also more reliable. However, you can use AI to acquire some basic knowledge and familiarize yourself with the topic." [R63]. Other respondents outline the need for proper education to guarantee accessibility from the use of GenAI tools "I think that if you are well informed about AI, you can avoid problems and hidden errors. If there were AI training courses or similar programs for everyone to bring them up to the same level of knowledge, that would be a great advantage." [R474]. The same respondent also points out inequalities in society that need to be tackled first in order to guarantee equal access to justice: "As soon as there are differences in education, i.e., large gaps in knowledge, then everything is no longer equal. The use of AI would be beneficial if everyone had access to this knowledge" [R474].

Yet, other respondents are sceptical about the utility of GenAI tools and rather fear a decrease of access to justice. Respondents point out problems like miscommunication ("If you express yourself too vaguely, the problem may not be clear. This can lead to major misunderstandings." [R266]), or disadvantages for specific population groups ("While AI can facilitate access to legal advice, there is a risk that it will be used as a substitute for professional advice, leaving users reliant on incomplete or incorrect advice, especially in complex cases where human expertise and contextual understanding are crucial. In addition, AI-driven processes could exacerbate unequal power relations, as less tech-savvy or informed users are disadvantaged and AI is unable to accurately capture the individual, often complex, and emotional nuances of a case. Given these risks, reliance on AI could lead to unfair outcomes and jeopardize the rights of individuals in important legal matters."[R372]).

*Case Complexity vs. Relief of the Legal System*

Respondents also referred to the case complexity as a central issue when calibrating their own risk acceptance of AI use for legal tasks. Respondents especially outline the potential for quicker solutions of legal conflicts as well as a general relief of the legal system through the use of AI. A respondent states: "In cases where the facts of the matter need to be assessed, AI can quickly remedy the situation and evaluate the situation itself. This can sometimes save costs and a great deal of time." [R279] These efficiency gains are considered important to relieve the legal system: "In times of increasing skills shortages, AI can provide useful support in reducing the workload of the courts by reaching satisfactory out-of-court settlements." [R226] Another respondent also highlighted those benefits as long as they are embedded in a strong legal framework: "With clear ethical and legal frameworks in place, many potential risks can be effectively mitigated." [R51] The tenant vs. landlord conflict described in the scenario was often treated as a low-risk scenario: "In simple cases, such as this rental dispute, I see advantages in out-of-court settlements." [R318]

Yet, respondents are concerned about the complexity and individuality that legal conflicts normally entail: "AI can only be used to obtain general information; it cannot be used to make specific decisions or judgments in advance, as these are individual and may therefore differ from one another." [R251] Respondents fear that AI tools are superficial and are not able to account for case-specific contextual factors like personal circumstances of the involved parties: "Many legal aspects are only touched upon by the AI tool. However, the AI tool does not address the actual case and all the facts. In addition, the AI tool lacks the so-called soft factors that are particularly important in legal matters, e.g., in criminal law." [R57] In a similar vein, a respondent pointed out the technical capabilities of AI systems: "The legal system is complicated and requires study. You shouldn't leave decisions that go beyond normal technical applications to chips." [R91] These concerns can be a decisive factor in determining the risk as too high in such cases: "AI learns from cases and regulations and attempts to force processes into these patterns. The evaluation of individual aspects that are absolutely case-specific and cannot be found in other cases, or only rarely, is virtually non-existent. And this is a disadvantage for the parties involved. Cost and time savings do not outweigh this disadvantage." [R120]

*Trust Calibration*

Trust is another recurrent theme in the answers of respondents. We refer to this theme as trust calibration as many respondents brainstorm about adequate levels of trust, i.e. how much trust can be adequately put into legal AI systems. As with the other themes, we detect different levels of trust in AI systems for legal cases. Some respondents' default is to put no trust whatsoever in AI for legal support: "Since AI is not always reliable, I would never rely on it in a legal dispute. I would always prefer to obtain information myself or ask someone who is familiar with the subject." [R324]. In a similar vein, another respondent raises the following issue: "I am not sure whether it is the right path we are taking, relying more and more on technology and thereby becoming more alienated from ourselves. We are increasingly losing the ability to trust our own intuition. Instead of self-determination, we are moving towards external determination." [R104]

Some respondents advocated a more nuanced trust calibration. They warned against over-trusting AI generated output while valuing procedural help and efficiency gains: "Risks could be overlooked if blind reliance is placed on the AI's statements. However, the AI's statements could be used as an aid to highlight overlooked points that could lead to arbitration." [R476]

## 7. Discussion

Integrating AI systems into critical societal domains, such as the legal sector, is a sensitive issue that presents both opportunities and risks. Appropriate risk assessment and management procedures are therefore becoming increasingly important. Although the EU has introduced legal obligations for AI developers to adhere to, we argue that these measures may not sufficiently protect citizens from the adverse consequences of AI tools, particularly since common risk assessment practices are expert-biased and often only require first-party auditing, i.e., auditing by the companies themselves (Hartmann et al., 2024; Reuel et al., 2025). Furthermore, there is a lack of empirical evidence centred on citizens' perspectives on the risks of AI, particularly regarding the identification of risk and benefit factors, as well as trade-off considerations in the risk management process. Third, understanding risk perceptions and how citizens assess trade-offs between risks and benefits is critical for the broader societal deployment and perceived legitimacy of AI-based legal solutions. Our empirical study, which used a representative sample of 488 German citizens, contributes to closing those gaps by providing valuable insights into how citizens approach using GenAI tools in the legal sector.

### 7.1. The Use of AI in the Legal Sector – A Matter of Complex Trade-off Decisions

By focusing on the risk-benefit perceptions of citizens, these findings contribute to the growing body of literature on the scientific and public discussions about the impact of AI on the legal domain (Dhungel & Beute, 2024; Helberger, 2025; Martinho, 2024; Socol De La Osa & Remolina, 2024). Our findings demonstrate that integrating AI into the legal domain is complex. Even if the judiciary or a government decides to rule out legal AI solutions, this is not yet a guarantee that citizens would find these useful or acceptable. The landlord vs. tenant scenario chosen for this study served as a prototype for a legal conflict in which AI could play a more prominent role in the near future. More specifically, we looked into the potential of GenAI to allow citizens to inform themselves about the law, and to enable alternative dispute resolution in the sense of mediation. This scenario was considered plausible with moderate risk severity and magnitude, meaning it is relatable and, while not extreme, a risk that should not be ignored.

In identifying and coding the risk factors, we revealed several interesting insights. Citizens may not be experts, but they are aware of both, risks and benefits of AI, whereby at least in our sample, the number of possible risks outweighed the number of benefits. This in itself does not say anything about whether or not they find certain AI solutions acceptable, but it does indicate that to make AI solutions for legal consultation or mitigation acceptable in the eyes of citizens, AI developers have a broad range of risks to consider and mitigate. Importantly, not all of these risks are amendable to technological fixes. A substantial number of risks are more complex and societal in nature. For example, the risk counts and trade-off explanations emphasize that citizens are primarily concerned about the role of humans, especially the preservation of human traits that make legal AI solutions acceptable such as emotions and empathy. When we compare our results with those of other empirical studies, we find some similarities. Mun, Yeong, et al. (2025) emphasize the importance of societal impact considerations and the values of care and fairness when evaluating the acceptability of AI use for legal decision-making. Further, German citizens highlight efficiency gains and relieving the burden on courts as central benefits of using GenAI in the legal system (Chien & Kim, 2024; Dhungel & Beute, 2024; Schwarcz et al., 2025). In terms of risk perception, German citizens are concerned that GenAI tools produce errors, and not without reason (Magesh et al., 2024). Furthermore, in line with surveys of judges, many respondents also express concerns about the human element and the context specificity of legal decision-making (Martinho, 2024; Socol De La Osa & Remolina, 2024). However, we also found some additional insights. For example, German citizens are concerned about manipulation potential and biases resulting from one-sided information provision. Furthermore, our results show that various dimensions of trust plays a central role in risk-benefit perceptions. Specifically, German citizens fear that people might overtrust AI in the legal domain, which could lead to detrimental personal (e.g., losing a court case) or societal consequences (e.g., erosion of trust in the legal system). These citizen-centered insights add valuable perspective to the risk-perception literature on AI in the legal field, which primarily focuses on the views of legal professionals.

Our thematic analysis of the trade-off explanations revealed that many citizens engage in complex trade-off decisions when considering risks, carefully navigating between potential gains and challenges. In particular, we identified several overarching themes in these explanations. However, even these themes are contested, ranging from an emphasis on risks to an emphasis on benefits. For example, while some respondents highlighted the potential for increased neutrality and fairness in decisions due to AI, others raised concerns about the potential for manipulation of AI systems. Another crucial dividing line was between access to justice and the accessibility or utility of AI. Some respondents expressed hope for greater equality due to decreasing costs, while others questioned this openness, highlighting potential hurdles for certain population groups. This indicates an important discussion that needs to be held in the academic field and in practice. Parties with more AI knowledge of how to use GenAI for their benefit will potentially achieve better outcomes from the use of GenAI for legal purposes, which might lead to a further amplification of societal differences.

Our results further enrich the literature on public perceptions of AI trade-offs (Kieslich et al., 2022; König et al., 2022) by showing which arguments are prioritized and *why*. These arguments are important to understand because they give indications on which specific concerns to practically address when developing risk mitigation strategies. They also demonstrate that currently, it is unlikely that a broader societal consensus can be achieved on the acceptability of legal AI tools. Concretely, this means that for now it is unlikely that legal solutions to provide people with access to legal information or dispute resolution should be replaced with AI, but rather be offered in parallel.

Additionally, our mixed-methods design provides insight into the predictors of risk acceptance in the legal domain. Consistent with the empirical findings of other studies (Marcinkowski et al., 2020; Mun, Au Yeong, et al., 2025; Shin et al., 2022), we also found that perceptions of AI unfairness are related to greater risk perceptions. Interestingly, however, the regression results did not yield significant findings for subjective knowledge of AI or personal usage of GenAI tools, adding further empirical evidence to related studies in the field of public opinion on AI and technology acceptance that show diverging effects of AI knowledge and usage on technology acceptance and risk perceptions (for non-significant associations see Araujo et al., 2020 and Wenzelburger et al., 2024; for significant association see Said et al., 2023). Moreover, trust in the legal system did not affect risk acceptance. This is an interesting finding since one could have argued that if people trust the legal system their risk acceptance is higher (e.g. because they have procedural rights and know decisions can be contested). These findings underscore the relevance of public attitudes as a *belief* in the capabilities of AI. This belief outweighs personal experience with and knowledge of these tools. Thus, in line with other researchers, we argue for centring research more around AI narratives and their connection to public imaginaries (Bareis & Katzenbach, 2022; Katzenbach, 2021; Sartori & Theodorou, 2022). It is important to point out that these beliefs can come with fundamental misconceptions about key capabilities but also limitations of AI. We saw significant divergences in the extent to which citizens were able to arrive at a realistic and nuanced assessment of individual and societal level consequences of the integration of AI into the legal system. That is, the open answers varied widely in argumentative depth with some answers mirroring common AI solutionist narratives while others engaged in nuanced trade-off discussion regarding risk and benefits for individuals and society. As citizens themselves indicated, differing levels in AI literacy, if not addressed, can ultimately result in significant divides in AI utility and, ultimately, lead to differences in procedural fairness and equality. This finding also has implications for public sphere actors, such as journalists, politicians, NGOs, and companies, as they are responsible for setting narratives aligned with the factual capabilities of AI. In our case, this means addressing the fairness of AI systems used in the legal domain to ensure the ways it is portrayed is well-calibrated to actual AI capabilities.

### 7.2. Implications for Risk Management

Our results show that citizens bring their own concerns and expectations to the table when it comes to deciding whether or not the deployment of legal AI tools is acceptable and provides utility. This means that both AI deployment but also risk management need to also navigate the lived experiences and realities of citizens. We argue that risk assessors must carefully consider these perceptions and incorporate them into their decision-making processes, for instance, when developing safeguards that address frequently mentioned concerns. Furthermore, our findings empirically support scholars' warnings that first-party audits tend to focus on the technical aspects of risks while ignoring societal concerns (Reuel et al., 2025). However, the DSA and the EU AI Act typically only mandate first-party audits, creating a structural problem in AI risk management. Interpreting this study in terms of participatory auditing (Hartmann et al., 2024), we stress that citizens showcase valuable concerns that might be overlooked in standard risk assessment procedures.

Our results show that many of the risk categories and trade-off explanations cannot be fixed technically. For instance, these issues include the role of humans and access to justice versus the accessibility of AI, which are challenges, that are also in conflict with the promised cost and efficiency gains highlighted by companies. These issues warrant broader discussions about equipping citizens with the means to engage with AI systems, safeguarding them against manipulation attempts, and ensuring human involvement in legal processes. The latter, in particular, raises questions about the risk classifications of AI systems in general. While the EU AI Act defines the use by judges for "administration of justice and democratic processes" as a high-risk case (Annex 3, EU AI Act), some respondents argue that it should be defined as an unacceptable risk. Citizens who argue this point say that AI can never replace human judges because it lacks empathy and cannot grasp the specific context of legal cases. Thus, regulators could consider these insights when negotiating a more detailed redefinition of AI risk categories for use in the legal sector.

### 7.3. Limitations

Our study has several limitations that need to be acknowledged. First, our study focuses on the perceptions of the German population. While Germany is a key player in the EU and adheres to EU regulations, it is still a single country with its own particular context. Cross-national studies involving Germany have shown that public opinion of AI varies among German citizens and citizens of other countries, including EU countries (Masso et al., 2023; Ullstein et al., 2025). Thus, we encourage studies that examine other countries' contexts, ideally using cross-national samples that enable researchers to compare identified risk factors and trade-off themes. Second, we only tested for two different legal tasks performed by generative AI systems. Yet, GenAI can be used for many more tasks throughout the legal process such as using GenAI for collecting evidence, drafting documents, or even for recommending sentences. Thus, our study only provides a snapshot of risks and benefits trade-offs associated with specific legal tasks. Future studies should extend our research on other legal AI tasks. Third, when conducting thematic analysis, we must address the issue of introducing our own expert bias[9] into the interpretation of data. We acknowledge that other research teams or even lay stakeholders may derive different themes from the qualitative analysis.

## 8. Conclusion

Artificial intelligence is witnessing a surge in applications within the legal sector. Citizens use GenAI as an accessible, inexpensive tool to inform themselves about legal options, while legal professionals use it to reduce their workload and speed up processes. To prevent harm to individuals, the legal system, and society as a whole, appropriate risk management processes are needed. In this paper, we argue that the current risk management process falls short of acknowledging and incorporating the perspectives of citizens, who are primarily affected by the technology. Thus, we set out to investigate these perspectives in terms of identifying risk factors and accepting risks, as well as highlighting concrete thoughts about benefit-risk considerations. Our results demonstrate that these insights are valuable and should be considered by regulators and companies.

**Conflicts of interest/Competing interests:** The authors declare no competing interests.

**Ethical approval and informed consent statements**: We received ethics approval from the institutional ethics board of the first authors' affiliation.

**Data availability statement**: The data and analysis script will be made open access at OSF upon acceptance in a peer-reviewed journal. While under review, files are available on request.

**Use of AI statement**: We used DeepL for text editing and translation of quotes. The translation of the quotes was checked by German native-speakers. We did not use GenAI for drafting any content of the paper nor for data analysis purposes.

---

[9] We want to acknowledge our positionality. Both of the authors who conducted the thematic analysis are native German speakers. Additionally, both hold a PhD in communication science. While the former enables us to understand the language and social context adequately, the latter may introduce a bias given the professional academic lens through which the authors view the subject.

**Appendix 1: Scenario Stimulus**

Julia had been having trouble with her apartment for several months. The issues started small - her shower seemed a little colder, but it grew progressively colder over the next few days. The first day, she had joked with a coworker that the cold water felt refreshing, but by the end of the week the disruption to her usual schedule had lost its charm. She contacted her landlord Alex, but initially received no response. And then on a weekend, things got quickly worse - the water suddenly became brown and dirty. The smell was terrible, too… and before long, the disgusting water had spread to her kitchen pipes as well.

Julia had been using generative AI to assist her in daily life and she wondered if it might help her with this issue as well. Using an online tool, she was able to describe the issues and draft a simple legal complaint. The AI also helped her document the issues -- by uploading pictures of the water and copies of her emails, the tool input those into her legal complaint as exhibits. The tool also directed her to her local small claims court, where she filed a complaint against her landlord. Although the 50 EUR filing fee was a pain, she knew she had saved the much greater cost of a lawyer. After the complaint was served, her landlord Alex received it with some concern. Alex acknowledged internally that he should have fixed the pipe issue sooner, but was reluctant to admit that -- what if it hurt his standing in the legal case? He consulted with a generative AI tool as well, and felt comfortable disclosing he felt he had made a mistake. The tool explained evidentiary rules prohibiting the admission of corrective measures in court, and gently encouraged him to make repairs, and perhaps even to seek some kind of mediation or conciliation with Julia. After also speaking with some friends, he did both. But Julia was upset and felt she would get a better result by going to court (and had already spent money on the filing fee), so the situation didn't resolve.

A few weeks later, the two arrived at a local civil court for a small claims hearing or trial. A mediator from the court greeted them, and guided them to courtroom tablets. "We strongly encourage parties to mediate their case, rather than to go to trial. In a mediation, you have a say in the outcome and can find a resolution you can accept. But in trial, the Court may order anything, or something you do not want to accept. The AI tool here can provide you with some information about the law. It cannot give you advice on your case, but it may help you understand your position. And, it may also ask you what you feel would be a fair outcome."

Julia and Alex each spent some time using the tablets and communicating with a generative AI tool. Although it didn't answer their questions perfectly, it did discuss the potential outcomes in the case. The tool alerted Alex that as a landlord, he could face certain extra damages for a failure to make timely repairs which gave him a greater concern for the case. And likewise, the tool reminded Julia that Alex had made repairs, even if a bit slowly, and noted that the lease had another 2 years so working to settle and improve their relationship moving forward would be in both parties interest. As the courtroom buzzed with other parties that were waiting to go to trial or have their small claims resolved, Alex and Julia were able to re-evaluate their cases with the AI tool. In doing so, the tool prompted each of them for terms for mediation they might accept, and worked through questions with them. By the time their case was called, the two had agreed through the tool to a mediation where Julia would receive one month free rent for her troubles. The tool drafted a short and plain statement for the two to review and sign, and they left the courtroom.

## Appendix 2: Codebook for Risks and Benefits

*Appendix 2a: Legal Consultation Risks*

| Code name | Description |
|---|---|
| Limited accessibility | Reference to the risks that AI might be not equally accessible for everybody, or that the access to it can implicate difficulties.<br>It also includes the following mentions:<br>• It is easier to communicate with humans<br>• Users need to have a minimum knowledge of AI tools to properly use it<br>• Insecurity of users |
| Over-trust | Reference to human over-trust in AI consultations. That includes, for instance:<br>• Blind trust in AI<br>• Taking recommendations with a grain of salt<br>• People show more trust in AI than humans<br>• Humans overestimating AI advice, and making bad decisions |
| Bias | References to biases of the AI system, especially referring to injustice, one-sidedness, or a lack of neutrality, for processing of data or the output of the system (e.g. consultations result, or verdict). This encompasses statements that mention prejudice, or non-objectivity/subjectivity of AI. It should also be coded if vulnerable groups (e.g. the elderly, migrants) are explicitly mentioned.<br><br>Yet, if the statements mostly refers to the system/knowledge base, code 8 (Incompleteness/manipulation of training data/knowledge base) should be coded. For instance, statements like "the system adheres to the programmer's input." |
| Complexity of cases | Reference to the complex nature of legal conflicts. Legal conflicts are perceived as too complex and individual as to be satisfactorily getting solved by an AI tool. |
| Completeness/Manipulation of information provision | Reference to the requirement of complete and/or correct input data by its users to show adequate performance. If information provision is flawed, bad or incomplete recommendations can be the result.<br>A specific notion here could also be the subjectivity of input, i.e. that an AI tool is only as good as the information provided. Information provision can be incomplete, or framed in favour of one party. (Thus, it only works if both parties are honest.). This code refers to the user's input, mostly described actively (e.g., "The AI only knows what information is typed in"). Also code, when manipulation concerns by users are mentioned, e.g. reference to a malicious intent of users to provide wrong information that push the case in their favour. |
| | Only code if there is an <u>explicit</u> reference to a user. If the description doesn't mention a user input, code as 8. |
| AI related errors | Reference to AI making mistakes in its output. It encompasses also the following:<br>• Incorrectness of output<br>• Vagueness of output<br>• Misleading presentation of information |
| Damage for legal professionals | Reference to negative consequences for legal professionals, i.e. job loss or the notion that lawyers / judges become irrelevant |
| Incompleteness/manipulation of training data/knowledge base | Reference to the concern that the legal AI tool works on insufficient training data, i.e. has insufficient legal knowledge (e.g., AI doesn't know the laws). This concern can be mentioned as a predecessor of concerns about insufficient output. Code also, if there is a mention of manipulation on the systemic level, e.g. programmers/AI companies code in their understanding of the law. It also includes statements that reference to a lack of transparency about how AI generates its recommendation. As such it hints to a concern about insufficient knowledge on how to evaluate the correctness of recommendation<br><br>In comparison to code 5 (completeness of information provision) this code refers to the sophistication of the model and focuses on the systems itself and/or its developers and deployers, and not the input that users provide. |
| Credibility of AI | Reference to the public acceptance of AI generated recommendations and decisions. For instance, it encompass the notion of who will take AI decisions seriously, or the concern that there is a lack of trust in the population towards AI use for legal purposes.<br><br>It also reference to perceptions that AI can never replace human legal professionals as there is no judicial basis present to do so. |
| Privacy and data protection | Reference to concerns about data that is provided to AI. Especially in case of personal legal cases it can encompass the fear that data might be unprotected, and/or being shared with third parties. |
| No human element | Reference to the notion that humans are important in the evaluation of legal conflicts. It encompasses notions like a lack of empathy, or missing humanness in recommendations and judgments. It also includes statements that humans are in general preferred to solve legal conflicts.<br><br>Note: Code loosely as far as humanness in form of human abilities, empathy, etc is described. |
| Real-world consequences of AI advices | Reference to the notion that AI recommendations and decisions can have real-world impacts. AI recommendations |

| | can be incorrect, inappropriate, or useless. Real-world consequences encompasses results like losing the specific court case. |
|---|---|
| Emotional responses to outcomes | Reference to negative emotional responses to the AI use. For instance, mentions that users are unhappy/unsatisfied with the AI recommendation or decision. |
| Fear of AI dependence | Reference to the concern that individuals / society might rely too much on AI and that the legal system is too dependent on technology (e.g. locked in). |
| Loss of trust in legal system*[10] | Reference to the concern that trust in the legal system (and actors) might get lost if people rely too much on AI. |
| Matureness of technology* | Reference to the concern that AI technology might not be advanced enough (yet) to offer reliable help. |
| Increase in law suits* | Reference to the concern that the lowered burden for AI recommendation can make it easier to file lawsuits and, thus, lead to an increase in law suits in general. |
| No possibility for follow-up interactions* | Reference to the concern that some tools might not allow a follow-up interaction which could lead to a lack of thoroughness in case assessment. |
| No AI required | Reference to a rejection of AI as it is generally perceived that AI is not required / shouldn't be used to solve a legal conflict. That includes mentions like that it is good to find agreement outside of courts, or that human conversations are preferable. |
| N.A. / Unspecific / No risks | Non-sensical answer, or explicit mention of don't know. It also includes answers that only relate to benefits or mention that there are no risks. Further, it includes answers that are too vague to be clearly assigned a code to. |

---

[10] Stars indicate that the code is unique to the condition

*Appendix 2b:* Legal Consultation Benefits

| Code name | Description |
| --- | --- |
| Low costs | Reference to the possibility of lower or no costs for a legal consult. This is often mentioned in terms of saving the costs for human lawyers (e.g. a fist consult) |
| Speed | Reference to the immediateness of legal consultation. Efficiency gains in the form of quicker procedures are expected. |
| Procedural help | Reference to multiple options of help during the legal process. This encompasses support in drafting documents, documenting cases, or writing letters. It also includes the expectation that AI can give a good and comprehensive overview about legal options.<br><br>It can also reference to the opportunity to get an easy first case assessment without contacting third parties. |
| Relieving the burden of legal professionals | Reference to the expectation that legal professionals will get unburdened. It also encompasses the notion that judges / attorney will have more time for serious cases. |
| Support for vulnerable populations / Accessibility | Reference to the hope that vulnerable groups might have easier access to justice (e.g. through lower cost tools, and/or translation options). AI can also simplify legal language to make cases more comprehensible. |
| Objectivity / Neutrality | Reference to the expectation that AI is perceived neutral, and can thus process data without bias and also come to a fairer decision than a human. It also encompasses statements that indicate that no human element might be a good thing (e.g. due to bias). |
| Less bureaucracy* | Reference to the expectation that legal conflict resolution can get less bureaucratic. |
| Lower fear of legal procedures* | Reference to the expectation that affected parties are less afraid to take part in a legal conflict / court case. |
| N.A. / No benefits / unspecific | Non-sensical answer, or explicit mention of don't know. It also includes answers that only relate to risks, or that there are no benefits. It also includes answers that are too vague to be clearly assigned a code to. |

*Appendix 2c:* Legal Mediation Risks

| Code name | Description |
|---|---|
| Limited accessibility | Reference to the risks that AI might be not equally accessible for everybody, or that the access to it can implicate difficulties.<br>It also includes the following mentions:<br>• It is easier to communicate with humans<br>• Users need to have a minimum knowledge of AI tools to properly use it<br>• Insecurity of users<br>• Human conversation preferable<br>• Face to face interactions get lost |
| Over-trust | Encompasses also the following:<br>• Blind trust in AI<br>• Taking recommendations with a grain of salt<br>• People show more trust in AI than humans |
| Bias | References to biases of the AI system, especially referring to injustice, one-sidedness, or a lack of neutrality, for processing of data or the output of the system (e.g. consultations result, or verdict). This encompasses statements that mention prejudice, or non-objectivity/subjectivity of AI. It should also be coded if vulnerable groups (e.g. the elderly, migrants) are explicitly mentioned. |
| Complexity of cases | Reference to the complex nature of legal conflicts. Legal conflicts are perceived as too complex and individual as to be satisfactorily getting solved by an AI tool. For this code it is essential that the legal case mentioned. |
| Completeness/Manipulation of information provision | Reference to the requirement of complete and/or correct input data by its users to show adequate performance. If information provision is flawed, bad or incomplete recommendations can be the result.<br>A specific notion here could also be the subjectivity of input, i.e. that an AI tool is only as good as the information provided. Information provision can be incomplete, or framed in favour of one party. (Thus, it only works if both parties are honest.). This code refers to the user's input, mostly described actively (e.g., "The AI only knows what information is typed in"). Also code, when manipulation concerns by users are mentioned, e.g. reference to a malicious intent of users to provide wrong information that push the case in their favour.<br><br>Only code if there is an <u>explicit</u> reference to a user. If the description doesn't mention a user input, code as 8. |
| AI related errors | Reference to AI making mistakes in its output. It encompasses also the following:<br>• Incorrectness of output<br>• Vagueness of output<br>Misleading presentation of information |
| Damage for legal professionals | Reference to negative consequences for legal professionals, i.e. job loss or the notion that lawyers / judges become irrelevant |

| Incompleteness/manipulation of training data/knowledge base | Reference to the concern that the legal AI tool works on insufficient training data, i.e. has insufficient legal knowledge (e.g., AI doesn't know the laws). This concern can be mentioned as a predecessor of concerns about insufficient output. Code also, if there is a mention of manipulation on the systemic level, e.g. programmers/AI companies code in their understanding of the law. It also includes statements that reference to a lack of transparency about how AI generates its recommendation. As such it hints to a concern about insufficient knowledge on how to evaluate the correctness of recommendation

In comparison to code 5 (completeness of information provision) this code refers to the sophistication of the model and focuses on the systems itself and/or its developers and deployers, and not the input that users provide. |
|---|---|
| Credibility of AI | Reference to the public acceptance of AI generated recommendations and decisions. For instance, it encompass the notion of who will take AI decisions seriously, or the concern that there is a lack of trust in the population towards AI use for legal purposes.

It also reference to perceptions that AI can never replace human legal professionals as there is no judicial basis present to do so. |
| Privacy and data protection | Reference to concerns about data that is provided to AI. Especially in case of personal legal cases it can encompass the fear that data might be unprotected, and/or being shared with third parties. |
| No human element | Reference to the notion that humans are important in the evaluation of legal conflicts. It encompasses notions like a lack of empathy, or missing humanness in recommendations and judgments. It also includes statements that humans are in general preferred to solve legal conflicts.

Note: Code loosely as far as humanness in form of human abilities, empathy, etc is described. |
| No judicial basis for AI use | Reference to the perceptions that AI can never replace human legal professionals as there is no judicial basis present to do so. |
| Real-world consequences of AI advices | Reference to the notion that AI recommendations and decisions can have real-world impacts. AI recommendations can be incorrect, inappropriate, or useless. Real-world consequences encompasses results like losing the specific court case, or the notion that one, or both parties could have reached a better result without AI. Given the stimulus, the real-world consequences might not be overly severe (e.g., no physical damage), but could rather express itself in financial losses etc. |
| Emotional responses to outcomes | Reference to negative emotional responses to the AI use. For instance, mentions that users are unhappy/unsatisfied with the AI recommendation or decision. |
| Fear of AI dependence | Reference to the concern that individuals / society might rely too much on AI and that the legal system is too dependent on technology (e.g. locked in). |
| No accountability* | Reference to the concern that accountability attribution might get difficult if AI systems are used for mediation (who is responsible, or can be contested for a (wrong) decision?) |
| No AI required | Reference to a rejection of AI as it is generally perceived that AI is not required / shouldn't be used to solve a legal conflict. That includes mentions like that it is good to find agreement outside of courts, or that human conversations are preferable |
| Reference to previous answer* | Reference to the previous answer. In this case, copy the code from the variable "Consultation risk" |
| N.A. / Unspecific / No risks | Non-sensical answer, or explicit mention of don't know. It also includes answers that only relate to benefits or mention that there are no risks.<br><br>Further, it includes answers that are too vague to be clearly assigned a code to. |

*Appendix 2d:* Legal Mediation Benefits

| Code name | Description |
| --- | --- |
| Low costs | Reference to the possibility of lower or no costs for a legal consult. |
| Speed | Reference to the immediateness of legal consultation. Efficiency gains in the form of quicker procedures are expected. |
| Procedural help | Reference to multiple options of help during mediation. This encompasses good and easy support in decision-making, good case support (e.g. through an overview of options), strengthening of negotiation positions, and improvement of legal speech. |
| Relieving the burden of legal professionals | Reference to the expectation that legal professionals will get unburdened. It also encompasses the notion that judges / attorney will have more time for serious cases. Further, it includes more general notions like it is good to reach an outer-court agreement. |
| Support for vulnerable populations / Accessibility | Reference to the hope that vulnerable groups might have easier access to justice (e.g. through lower cost tools, and/or translation options). AI can also simplify legal language to make cases more comprehensible. |
| Objectivity / Neutrality | Reference to the expectation that AI is perceived neutral, and can thus process data without bias and also come to a fairer decision than a human. It also encompasses statements that indicate that no human element might be a good thing (e.g. due to bias). |
| Good but only with human oversight* | Reference to the notion that AI for mediation is general positive, but only if humans are kept in the loop. |
| Help in building human connection* | Reference to the expectation that negotiating with the support of an AI tool can enhance communication between the human parties. |
| Lower emotional damage* | Reference to the expectation that AI mediated mediations inflict less emotional damage for the parties than appearing before a (human) court. |
| Less need for legal professionals* | Reference to the notion that there is less need for legal professionals, which is perceived as a good thing (e.g., because of mistrust in legal professionals). |
| No difference to scenario before / Reference to previous answer* | Reference to the previous answer. In this case, copy the code from the variable "Consultation benefit" |
| N.A. / No benefits / unspecific | Non-sensical answer, or explicit mention of don't know. It also includes answers that only relate to risks, or that there are no benefits. It also includes answers that are too vague to be clearly assigned a code to. |